\documentclass[11pt,a4wide]{article}
\pdfoutput=1
\usepackage{jheppub}
 \usepackage{graphicx}
 \usepackage{bm}
 \usepackage{braket}
\usepackage{epsf,
shadow,pifont}
 \usepackage{amsmath,epsfig}
 \usepackage{amssymb,amsfonts,amsthm,amstext,amscd,dsfont}
\usepackage{latexsym}
\usepackage{slashed}
\usepackage{float}
\usepackage{epsfig}
\usepackage{simplewick}

\newcommand{\be}{\begin{equation}}
\newcommand{\ee}{\end{equation}}
\newcommand{\ba}{\begin{eqnarray}}
\newcommand{\ea}{\end{eqnarray}}

\newcommand{\nn}{\nonumber\\}

\def\pa{\partial}

\def\a{\alpha}

\def\g{\gamma}
\def\G{\Gamma}

\def\D{\Delta}

\def\vp{\varphi}

\def\t{\tau}

\def\lra{\leftrightarrow}

\title{Closed-form expression for cross-channel conformal blocks near the lightcone}

\author{Wenliang Li}
\affiliation{Okinawa Institute of Science and Technology Graduate University, 1919-1 Tancha, Onna-son, Okinawa 904-0495, Japan}

\emailAdd{lii.wenliang@gmail.com}

\abstract
{In the study of conformal field theories, 
conformal blocks in the lightcone limit are fundamental to the analytic conformal bootstrap method. 
Here we consider the lightcone limit of 4-point functions of generic scalar primaries. 
Based on the nonperturbative pole structure in spin of Lorentzian inversion, 
we propose the natural basis functions for cross-channel conformal blocks. 
In this new basis, we find a closed-form expression for crossed conformal blocks in terms of the Kamp\'e de F\'eriet function, 
which applies to intermediate operators of arbitrary spin in general dimensions.  
We derive the general Lorentzian inversion 
for the case of identical external scaling dimensions. 
Our results for the lightcone limit also shed light on the complete analytic structure of conformal blocks in the lightcone expansion. 
}

\begin{document}

\maketitle

\section{Introduction}
Many important physical systems are described by conformal field theories (CFTs), ranging from statistical physics to high energy physics. 
The conformal bootstrap program aims at classifying and solving conformal field theories using general principles \cite{Ferrara:1973yt, Polyakov:1974gs}, 
such as conformal invariance, crossing symmetry, unitarity and analyticity.  
Although considerable results in 2d have been obtained for a long time \cite{DiFrancesco:1997nk}, 
the conformal bootstrap in higher dimensions have made major progress only since the work of \cite{Rattazzi:2008pe}. 
This modern numerical approach of the conformal bootstrap has led to nontrivial bounds on the parameter space of unitary CFTs. 
The impressive results include the precise determinations of the 3d Ising critical exponents 
\cite{ElShowk:2012ht, El-Showk:2014dwa, Kos:2014bka,Kos:2016ysd}. 
We refer to \cite{Poland:2018epd} for a comprehensive review. 

In parallel, the analytic approach has also made notable progress after the revival of the $d>2$ conformal bootstrap. 
By considering the lightcone limit of the crossing equation, 
it was shown in \cite{Fitzpatrick:2012yx, Komargodski:2012ek} that the twist spectrum of  a $d>2$ CFT is additive at large spin: 
if the twist spectrum contains scalars of twist $\t_1, \t_2$, 
then there will be accumulation points at $\t_1+\t_2+2n$ with $n=0,1,2,\dotsb$ for the large spin sector. 
An earlier discussion in a more specific context can be found in \cite{Alday:2007mf}.
The analytic conformal bootstrap can be formulated as an algebraic problem \cite{Alday:2015ewa, Alday:2016njk, Alday:2016jfr}. 
More recently, significant advances towards the nonperturbative regime have been made 
by upgrading the analytical toolkit 
from asymptotic expansion at large spin 
\cite{Alday:2015ewa, Alday:2016njk, Alday:2016jfr, Alday:2015eya, Kaviraj:2015cxa, Alday:2015ota, Simmons-Duffin:2016wlq, Dey:2017fab, Cardona:2018dov, Sleight:2018epi}
to convergent Lorentzian inversion at finite spin \cite{Liu:2018jhs,Cardona:2018qrt}.
\footnote{See \cite{Sleight:2018ryu} for convergent results from a different method.} 
They were based on the elegant Lorentzian OPE inversion formula proposed by Caron-Huot in \cite{Caron-Huot:2017vep} 
\footnote{See \cite{Simmons-Duffin:2017nub} for an alternate derivation and \cite{Kravchuk:2018htv}  for a generalization.}, 
which established the analyticity in spin assumed earlier \cite{Costa:2012cb}. 
As we will see, the Lorentzian inversion formula also provides new insights into the analytic expression of conformal blocks. 

We will consider 4-point functions of spin-0 primaries: 
\be
\langle \vp_1\,\vp_2\,\vp_3\,\vp_4\rangle
=
\bigg(\frac {x_{24}}{x_{14}}\bigg)^{2a}\,
\bigg(\frac {x_{14}}{x_{13}}\bigg)^{2b}\,
\frac {\mathcal G(z,\bar z)} {x_{12}^{\D_1+\D_2}\, x_{34}^{\D_3+\D_4}}\,,
\label{4-p-fn}
\ee
where $x_i$ denotes the positions of the external scalars $\vp_i$, 
$x_{ij}=|x_i-x_j|$ is the distance between two operators, 
$a=(\D_1-\D_2)/2, b=(\D_3-\D_4)/2$ are the differences in the external scaling dimensions, 
$z, \bar z$ are related to the conformally invariant cross-ratios by
\be
z\bar z=\frac {x_{12}^2\, x_{34}^2}  {x_{13}^2\, x_{24}^2},\quad (1-z)(1-\bar z)=\frac {x_{14}^2\, x_{23}^2}  {x_{13}^2\, x_{24}^2}\,.
\ee

In the conformal bootstrap method, the nontrivial equation from crossing symmetry schematically reads:
\be
\sum_{\mathcal O} {\text{direct conformal block}} =\sum_{\mathcal O} {\text{crossed conformal block}}\,,
\label{crossing}
\ee
where $\mathcal O$ indicates intermediate primary operators, and ``direct" and ``crossed" are short for direct- and cross-channels, 
sometimes called s- and t- channels. 
The two channels of operator product expansions (OPEs) are related by the crossing transform, 
which exchanges $\vp_1$ and $\vp_3$. 
The crossing equation \eqref{crossing} is a consequence of OPE associativity, namely correlation functions are independent of the order of operator product expansions. 
By solving \eqref{crossing} near the lightcone, 
the scaling dimensions and OPE coefficients of low-twist operators in the direct-channel are associated with 
the Lorentzian inversion of cross-channel conformal blocks in the lightcone limit. 

A conformal block encodes the contributions of a primary and its descendants, labeled by the scaling dimension and spin of the primary. 
As functions of two variables, conformal blocks satisfy the Casimir differential equations \cite{Dolan:2003hv, Dolan:2011dv}. 
In the lightcone limit, the quadratic equation for direct-channel conformal blocks is truncated to a closed equation with one variable. 
The solutions are the $SL(2,\mathbb R)$ conformal blocks, given by ${}_2F_1$ hypergeometric functions. 
On the other hand, 
the quadratic differential equation for generic cross-channel conformal blocks is not truncated to a closed equation. 
Accordingly, their analytic expressions are more sophisticated than those in the direct-channel.   

The direct-channel OPE of a scalar 4-point function gives
\be
\mathcal G(z,\bar z)=\sum_{i}\,P_i \, \tilde G_{\t_i,\ell_i}(z,\bar z)\,,
\ee
where $P_i$ is the product of two OPE coefficients associated with the intermediate operator $\mathcal O_i$, 
$\tilde G_{\t_i,\ell_i}(z,\bar z)$ denotes the direct-channel conformal blocks, 
$\t=\D-\ell$ is known as twist, 
and $\D_i, \ell_i$ are the scaling dimension and spin of $\mathcal O_i$. 

In this paper, we are interested in the lightcone limit of cross-channel conformal blocks:   
\be
G_{\t_i,\ell_i}(z,\bar z)=\tilde G_{\t_i,\ell_i}(1-z,1-\bar z)\Big|_{z\rightarrow 0}\,.
\ee
The direct and crossed blocks are related by the crossing transform 
$(z, \bar z)\rightarrow (1-z,1-\bar z)$. 
Our normalization is $G_{\t,\ell}(z,\bar z)\big|_{a=b=0}=-\,(1-\bar z)^{\t/2}\,\log z+\dotsb$ as $\bar z\rightarrow 1$. 

In general dimensions, the conformal block of an intermediate scalar can be written as Appell's function $F_4$ \cite{Dolan:2000ut}. 
One can easily derive the lightcone limit:  
\be
G_{\D,0}(z,\bar z)=
\frac{(1-\bar z)^{\D/2}}{(z\bar z)^{\frac {b-a} 2}}
\bigg(
\frac{\G(a-b)\,(z\bar z)^{\frac{b-a}2}}
{( \D/ 2)_{a}\,(\D/ 2)_{-b}}\,
{}_2F_1
  \bigg[
\begin{matrix}
  \D/2-a , \, \D/ 2+b\\
 \D-d/ 2+1
\end{matrix}\,;
1-\bar z
\bigg]
+(a\lra b)
\bigg)\,,
\label{spin-0}
\ee
where
\be
(X)_Y=\frac{\G(X+Y)}{\G(X)}\,,
\ee 
and $\G(X)$ is the gamma function.  

For spinning intermediate operators, compact expressions for the full conformal blocks were only known in 
$d=2,4,6$ dimensions \cite{Dolan:2000ut}\cite{Dolan:2003hv}. 
Here we are interested in the analytical expression in any dimensions. 
For a given spin, one can compute the crossed block using Dolan-Osborn's recursion relation in spin  \cite{Dolan:2000ut}, 
which transforms the spinning block into a linear combination of spin-0 blocks with polynomial insertions. 
Alternatively, the crossed spinning block can be calculated using the Mellin-Barnes representation. 
The lightcone limit again reduces to a linear combination of ${}_2F_1$ hypergeometric functions,  
where the coefficients are given by sums of ${}_4F_3$ hypergeometric series from Mack polynomials \cite{Mack:2009mi} \cite{Dolan:2011dv}. 
For identical external scaling dimensions, 
several leading series coefficients with arbitrary spin have been computed 
using recursion relations
\cite{Alday:2015ewa} 
or a quartic differential equation \cite{Caron-Huot:2017vep}, 
but no closed-form results were presented in \cite{Alday:2015ewa, Caron-Huot:2017vep}.

The cross-channel conformal blocks in the lightcone limit play a central role in the analytic conformal bootstrap method. 
However, the formulae from spin recursion or Mack polynomials are not satisfactory. 
They are a bit bulky, and hence the Lorentzian inversion and the associated numerical evaluation can be quite involved. 
It is especially hard to study the leading corrections to OPE coefficients, due to the complexity of the regular part in these formulae. 
Some results from Mack polynomials can be found in \cite{Cardona:2018qrt}.

In this paper, we will present a new closed-form expression \eqref{closed-form},
which naturally extends the spin-0 expression \eqref{spin-0} to the spinning case. 
For identical external scaling dimensions, 
we will also show in \eqref{general-LI} that the Lorentzian inversion of both the logarithmic and regular parts can be readily derived. 

\section{Nonperturbative poles in spin}
To find the natural basis functions, we will study the pole structure of the Lorentzian inversion of crossed conformal blocks. 
Since Riemann, the ${}_2F_1$ hypergeometric function is known to be characterized by 
the singularity structure of the hypergeometric differential equation, 
which explains its wide application, including the direct blocks near the lightcone. 
As a quadratic differential equation for the crossed blocks is not yet available 
\footnote{One can consider differential equations of higher orders \cite{Caron-Huot:2017vep}. 
For instance, closed-form results for the diagonal limit of conformal blocks were obtained using 
both quadratic and quartic Casimir equations \cite{Hogervorst:2013kva}.}, 
we will instead study the singularity structure of a nontrivial integral transform, the Lorentzian inversion. 
\footnote{Mack polynomials are associated with the Mellin transform.}

We will concentrate on the case of identical external scaling dimensions, 
which simplifies the analysis but still captures the main insights. 
In the lightcone limit, Caron-Huot's Lorentzian inversion formula 
reduces to an inversion formula for $SL(2, \mathbb R)$ conformal blocks \cite{Caron-Huot:2017vep,Alday:2017zzv}:
\be
{\bf L}[f(\bar z)]=\frac{(2h-1)\,\G(h)^2}{\pi^2\,\G(2h)}
\int_0^1 \frac{d\bar z }{\bar z^{2}}\,\bar z^h\,
{}_2F_1\big[
h,h, 2h;\, \bar z\big]
\text{dDisc}[f(\bar z)]\,,
\ee
where the double discontinuity is defined by analytic continuations around $\bar z=1$:
\be
\text{dDisc}\left[f(\bar z)\right]=f(\bar z)-\frac 1 2 f^\circlearrowleft(\bar z)- \frac 1 2 f^\circlearrowright(\bar z)\,.
\ee
Here $h=\t/2+J$, and $J$ indicates the spin of direct-channel operators. 
The eigenvalue of the quadratic $SL(2,\mathbb R)$ Casimir is $h(h-1)$. 
The Lorentzian inversion of a power-function building block is
\be
{\bf L}\left[\frac{(1-\bar z)^p}{\bar z^q}\right]
=\frac{2(2h-1)\,\G(h)^2\,\G(h-q-1)}{\G(-p)^2\,\G(p+1)\,\G(2h)\,\G(h+p-q)}\,
{}_3F_2\left[
\begin{matrix}
h,h,h-q-1\\
2h, h+p-q
\end{matrix};\,1
\right]\,.
\label{LI-bb}
\ee
The standard expansion around $\bar z=1$ uses $(1-\bar z)^p$ as the basis functions. 
Another convenient choice is $(1-\bar z)^p/\bar z^p$, 
since \eqref{LI-bb} becomes a product of gamma functions. 

It is straightforward to analytically continue \eqref{LI-bb} to small or negative $h$. 
Generically, there are poles at $h=q+1-k$ with $k=0,1,2,\dotsb$ due to $\G(h-q-1)$.  
As $q$ characterizes the singularity at $\bar z=0$, 
they are nonperturbative poles for the expansion around $\bar z=1$. 
The first of them is at $h=q+1$. 
The basis functions $(1-\bar z)^p/\bar z^p$ lead to an asymptotic series, 
as the first pole moves to larger $h$ at higher order. 
There are also poles at $h=-k$ from $\G(h)$. 
Some poles may disappear for certain values of $p, q$.

Perturbatively, the pole structure seems to depend on 
the choice of basis functions, 
but the nonperturbative results should be independent of our choice. 
As spurious poles will cancel out and nonperturbative poles will appear in the final results, 
it is reasonable to consider basis functions that are manifestly consistent with the nonperturbative pole structure, 
which should lead to a more natural formulation of conformal blocks.

\begin{figure}[h!]
\begin{center}
\includegraphics[width=10cm]{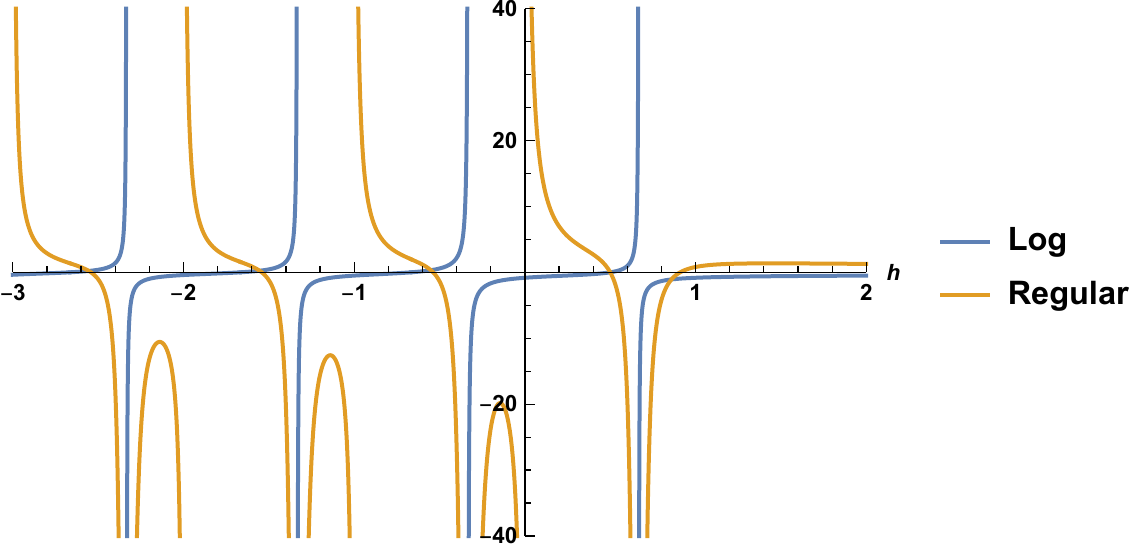}
\caption{Lorentzian inversion of the logarithmic (blue) and regular (orange) terms of the cross-channel conformal block, as functions of $h=\t/2+J$, where $\t, J$ are the direct-channel twist and spin.  
The spacetime dimension is $d=3$. The scaling dimensions of the intermediate and external scalars are $\D=1, \D_\vp=1/2+1/3$. 
There are poles at $h=2/3-k, -k$ with $k=0,1,2,\dotsb$. The first pole is at $2/3$, in accordance with \eqref{Pole-location}. 
The logarithmic case has only the first type of poles due to $q=p$. } 
\label{figure1}
\end{center}
\end{figure}

As an example of the pole structure, let us consider the intermediate scalar with $\D=d-2$. 
The crossed lightcone block is particularly simple:
\be
G_{d-2,0}(z,\bar z)=-\left(\frac{1-\bar z}{\bar z}\right)^{\frac d 2-1}
\Big(\log z-\log \bar z+2H_{\frac {d} 2-2}\Big)\,.
\label{spin-0-simple}
\ee
In the crossing equation, the cross-channel block is multiplied by $(1-\bar z)^{-\D_\vp}/\bar z^{-\D_\vp}$, 
so the first nonperturbative pole is at 
\be
\frac d 2-\D_\vp=1-\g_\vp\,,
\label{Pole-location}
\ee 
where $\D_\vp,\g_\vp$ are the scaling and anomalous dimensions of $\vp$. 
The pole positions of the $\log z$ term can be deduced from \eqref{LI-bb} by setting $p=q=(d-2)/2-\D_\vp$. 
For the regular part, one can generate the $\log \bar z $ term by taking a $q$-derivative. 
The concrete results of $d=3$, $\D=1$, $\D_\vp=1/2+1/3$ are presented in Fig. \ref{figure1}. 
Analytic continuation across the first pole was first discussed in concrete models in \cite{Alday:2017zzv}, 
which explicitly extends analyticity in spin to $J=0$ ! 
Our pole analysis is partly motivated by the remarkable results in \cite{Alday:2017zzv}. 

Then we consider spinning intermediate operators. 
There are additional singularities at $h=1-\D_\vp-k$, associated with the standard expansion. 
Surprisingly, although the residues change, the locations of nonperturbative poles do not vary with the spin and twist of $\mathcal O_i$, 
which hints at a hidden universal structure! 
Accordingly, the natural basis functions are
\be
\frac{(1-\bar z)^p}{\bar z^{d/2-1}}\,,
\label{basis-fun}
\ee
as they are manifestly compatible with the locations of nonperturbative poles. 
For generic external scalars, 
the exponents also depend on the external dimensions. 
Below, we will discuss the series expansion in terms of \eqref{basis-fun}. 

\section{Lightcone limit of crossed conformal blocks}
To see the general pattern, 
we consider several 3d examples with identical external scalars at low spin. 
They are computed using Dolan-Osborn's recursion relation in spin \cite{Dolan:2000ut} 
and Mack polynomials \cite{Mack:2009mi}\cite{Dolan:2011dv}, which lead to equivalent results. 
Then we expand the lightcone limit of crossed blocks in terms of \eqref{basis-fun}, and simplify the series carefully. 
As expected, they exhibit a universal pattern. 
The results can be further extended to arbitrary spacetime dimensions and generic external scalars. 
In the end, we arrive at a general expression in closed form: 
\ba
G_{\t,\ell}(z,\bar z)&=&
\frac{(1-\bar z)^{\t/2}\,(z\bar z)^{\frac {a-b} 2}}{\bar z^{d/2-1}}
\Bigg(
\frac{\G(b-a)\,(z/\bar z)^{\frac{a-b}2}}
{( \t/ 2+\ell)_{-a}\,(\t/ 2+\ell)_b}\,
\nn&&\qquad\qquad\times\,
F^{0,2,2}_{0,2,1}
  \bigg[
\Big|
\begin{matrix}
 -\ell,\, 3-d-\ell \\
\g, \,2- d/ 2 -\ell
\end{matrix}\,
\Big|
\begin{matrix}
  \g/2-a , \, \g/ 2+b\\
 \t/2+\g/ 2+\ell
\end{matrix}\,\Big| 
x,-x
\bigg]
+ \,(a\lra b)
\Bigg)\,,\quad
\label{closed-form}
\ea
where 
\be
x=1-\bar z\,,\quad
\g=\t-d+2\,.
\ee 
Note that $\g$ is the anomalous dimension when $\ell>0$. 
The Kamp\'e de F\'eriet function $F^{a_1,a_2,a_3}_{b_1,b_2,b_3}$ is a two-variable hypergeometric function. 
Our definition is
\begin{eqnarray}
&&F^{0,2,2}_{0,2,1}
\bigg[\Big|
\begin{matrix}
A_1, A_2\\
B_1,B_2
\end{matrix}
\Big|
\begin{matrix}
A_3, A_4\\
B_3
\end{matrix}
\Big|
x, y
\bigg]
=\sum_{m,n=0}^\infty\,
 \frac{(A_1)_n\,(A_2)_n\,}{(B_1)_n\,(B_2)_n}\,
  \frac{(A_3)_{m+n}\,(A_4)_{m+n}}{(B_3)_{m+n}}\,
  \frac{x^m\,y^n}{m!\,n!}\,.
 \qquad
\end{eqnarray}
The complete crossed blocks contain higher order terms in $z$. 

For $\ell=0$, our expression \eqref{closed-form} reduces to \eqref{spin-0} after a linear transformation of ${}_2F_1$. 
Although \eqref{spin-0} looks simpler, 
its direct generalization to $\ell>0$ will have much 
more complicated coefficients when expanded into ${}_2F_1$. 
The additional factor  $\bar z^{-(d/2-1)}$ from the nonperturbative-pole analysis 
is crucial to the compactness of \eqref{closed-form}. 

For non-negative integral spin
\footnote{One should set $\ell$ to a non-negative integer before taking the limit $d=2,4,6$. 
If we take the latter limit first, 
the results are different from those from the closed-form expressions in \cite{Dolan:2000ut}\cite{Dolan:2003hv}, 
but the differences can be obtained from \eqref{closed-form} by 
the spin-shadow transform $\D\rightarrow \D,\,\ell\rightarrow 2-d-\ell$. 
\label{footnote-even-d}}, 
our formula \eqref{closed-form} is consistent with the closed-form expressions in $d=2, 4, 6$ dimensions \cite{Dolan:2000ut}\cite{Dolan:2003hv}, 
and reproduces the leading series coefficients in \cite{Alday:2015ewa, Caron-Huot:2017vep}.
\footnote{We would like to point out some typos in the references. 
In the last line of eq. (2.20) in \cite{Dolan:2003hv}, 
the denominator should contain $(\D-\ell-4)(\D-\ell-6)$, rather than $(\D+\ell-4)(\D+\ell-6)$.
In the last line of eq. (A.35) in \cite{Caron-Huot:2017vep}, 
the second term of the numerator should be $\D(3-2J+2J^2)$, without a factor $2$.} 
Furthermore, \eqref{closed-form} is significantly simpler than the expressions in general dimensions 
from either Dolan-Osborn's recursion relation or Mack polynomials.

\section{Identical external scaling dimensions}
Using the $d=2,4$ analytical expressions in \cite{Dolan:2000ut}, closed-form results 
for the Lorentzian inversion of cross-channel conformal blocks 
were presented in \cite{Liu:2018jhs}.
\footnote{For generic spin $\ell$, when $0\ll z\ll \bar z\ll1$, the boundary condition in \cite{Caron-Huot:2017vep} is  
$\tilde G_{\t,\ell}(z,\bar z)\sim z^{\t/2}\, \bar z^{\t/2+\ell}(1+\dotsb)$, 
where $\dotsb$ indicates integer powers of $z/\bar z$ and $\bar z$. 
We use the same boundary condition. 
Note that the closed-form expressions in \cite{Dolan:2000ut}\cite{Dolan:2003hv}
behave differently as  
$\tilde G_{\t,\ell}(z,\bar z)\sim z^{\t/2}\,\bar z^{\t/2+\ell}[(1+\dotsb)+c_d(\ell)\,(z/\bar z)^{\ell+d/2-1}(1+\dotsb)]$.} 
Here, we are again interested in the lightcone limit in general spacetime dimensions and consider identical external scaling dimensions. 
The associated crossing equation is the first nontrivial equation for the analytic conformal bootstrap,  
the main equation in many concrete studies.

For identical scaling dimensions, the cross-channel conformal block \eqref{closed-form} becomes
\ba
G_{\t,\ell}(z,\bar z)&=&
-\big(\log z+2 H_{\frac \t 2+\ell-1}+\pa_\a\big)
\nn&&\times
\Bigg(\frac{(1-\bar z)^{\t/2}}{\bar z^{d/2-1+\a}}\,
F^{0,2,2}_{0,2,1}
  \bigg[
\Big|
\begin{matrix}
 -\ell,\, 3-d-\ell \\
\g, \,2- d/ 2 -\ell
\end{matrix}\,
\Big|
\begin{matrix}
  \g/2 -\a, \, \g/ 2-\a\\
 \t/2+\g/ 2+\ell
\end{matrix}\,\Big| 
x,-x
\bigg]
\Bigg)\Bigg|_{\a\rightarrow 0}\,,\quad
\label{KdF-identical}
\ea
where $H_X=H(X)$ is the harmonic number. 
In the limit
\be
\a=\frac{\D_1-\D_2} 2=\frac {\D_4-\D_3} 2
\rightarrow 0\,,
\ee 
the external scaling dimensions become identical. 
More explicitly, we can write \eqref{KdF-identical} as: 
\begin{eqnarray}
G_{\t,\ell}(z,\bar z)=-\left(\log z+2H_{\frac \t 2+\ell-1}+\pa_\a\right)
\sum_{n=0}^\ell\,
c_n\,g_n(\bar z)\,
\Big|_{\a\rightarrow 0}\,,
\label{general-expression}
\end{eqnarray}
where 
\be
c_n=\frac {\ell!}{n!\,(\ell-n)!}\,\frac{(3-d-\ell)_n\,\left[({\g}/ 2-\a)_n\right]^2}
{(\g)_n\,(2- d/ 2-\ell)_n\,(\t/2+\g/ 2+\ell)_n}\,,
\label{cn}
\ee
and
\be
g_n(\bar z)=\frac{(1-\bar z)^{\t/2+n}}{\bar z^{d/2-1+\a}}\,
{}_2F_1\bigg[
\begin{matrix}
{\g} /2+n-\a, {\g}/ 2+n-\a\\
\t/2+\g/ 2+\ell+n
\end{matrix};\,
x
\bigg]\,.
\label{gn}
\ee 
Note that the even $d$ limit and non-negative integer $\ell$ limit of $c_n$ do not commute 
because of the singularities from the denominator $(2- d/ 2-\ell)_n$. 

For a conserved current, the twist is $\t=d-2$ and the anomalous dimension $\g$ vanishes. 
We can evaluate the $\a$-derivative:
\begin{eqnarray}
G_{d-2,\ell}(z,\bar z)&=&-\frac{(1-\bar z)^{\t/2}}{\bar z^{\t/2}}\Bigg(\log z-\log\bar z+2\,H_{\frac \t 2+\ell-1}
\nn&&\qquad\qquad\quad
-\sum_{n=1}^\ell\,
\frac{(-\ell)_n\,(1-\t-\ell)_n\,(-x)^n}
{n\,(1-\t/ 2-\ell)_n\,( \t/ 2+\ell)_n}\,
{}_2F_1\bigg[
\begin{matrix}
n, n\\
\t/ 2+\ell+n
\end{matrix};\,
x
\bigg]
\Bigg)
\,.\qquad
\label{conserved}
\end{eqnarray} 
The scalar expression \eqref{spin-0-simple} is a special case of \eqref{conserved} with $\ell=0$. 
The global symmetry current corresponds to the case of $\ell=1$. 
Using a contiguous relation, 
the stress-tensor block also takes a compact form
\begin{eqnarray}
G_{d-2,2}(z,\bar z)&=&
-\frac{(1-\bar z)^{\t/2}}{\bar z^{\t/2}}\,\Bigg(\log z-\log\bar z+2H_{\frac \t 2+1}
-\frac{8\,(\t+1)\,x}{(\t+2)(\t+4)}
{}_2F_1\bigg[
\begin{matrix}
1, 2\\
\t/ 2+3
\end{matrix};\,
x
\bigg]
\Bigg)\,.\qquad
\end{eqnarray}

To perform the Lorentzian inversion, let us introduce $y=(1-\bar z)/\bar z$, 
then $g_n(\bar z)$ becomes a 0-balanced ${}_2F_1$ function: 
\be
g_n(\bar z)=y^{\t/2+n}\,
{}_2F_1\left[
\begin{matrix}
{\t}/ 2+\ell+\a, {\g}/ 2+n-\a\\
\t/2+\g/ 2+\ell+n
\end{matrix};\,
-y
\right]\,.
\label{gn-y}
\ee
Using the Mellin-Barnes representation \cite{Liu:2018jhs,Cardona:2018qrt,Sleight:2018ryu}, 
one can compute the Lorentzian inversion:
\begin{eqnarray}
{\bf L}\left[y^{-\D_\vp}g_n(\bar z)\right]
&=&
2(2h-1)\,\G(h-a_1+a_3)\,\G(h-a_1+a_4)\,\G(h)^2\,\frac{\G(a_1+a_2)}{\G(1-a_1)^2}\,
\nn&&\times\,
\psi(h+a_1+a_2-1 ; h,a_1,a_2,a_3,a_4)\,,
\label{gn-inversion}
\end{eqnarray}
where
\begin{eqnarray}
a_1=\frac \t 2+n-\D_\vp+1\,,&\quad&
a_2=\frac \g 2+\ell+\D_\vp-1\,,\nn
a_3=\frac \t 2+\ell+\a\,,&\quad&
a_4=\frac \g 2+n-\a\,.
\end{eqnarray}
We have used a very well-poised hypergeometric series: 
\be
\psi(A;B_1,B_2,B_3,B_4,B_5)
=\frac{\G(A+1)\, {}_7F_6\Bigg[
\begin{matrix}
A, A/2+1, B_1,\dots, B_5\\
A/2, A-B_1+1,\dots, A-B_5+1
\end{matrix}; \,1
\Bigg]}{\G\big(2A+2-\sum_{k=1}^5\,B_k\big)\prod_{k=1}^5\,\G(A-B_k+1)}\,,
\ee
which is associated with the Wilson function \cite{Wilson, Groenevelt}. 
According to $\G(h-a_1+a_4)$, the nonperturbative pole structure is manifest for the Lorentzian inversion of each $g_n(\bar z)$. 

Finally, the Lorentzian inversion of the crossed blocks can be readily derived 
from \eqref{general-expression} and \eqref{gn-inversion}:
\be
{\bf L}\left[y^{-\D_\vp}G_{\t,\ell}(z,\bar z)\right]=-\big(\log z+2H_{\frac \t 2+\ell-1}+\pa_\a\big)\,
\sum_{n=0}^\ell\,c_n\,
{\bf L}\left[y^{-\D_\vp}g_n(\bar z)\right]
\Big|_{\a\rightarrow 0}\,,
\label{general-LI}
\ee
which can be expressed in terms of the Kamp\'e de F\'eriet function. 
\footnote{The Lorentzian inversion of cross-channel conformal blocks can also be viewed as the crossing kernels or 6j symbols for the conformal group \cite{Liu:2018jhs}. The Wilson function already appeared in the study of crossing kernels in \cite{Hogervorst:2017sfd}. 
In \cite{Sleight:2018ryu}, the corrections to the anomalous dimensions of leading twist operators from a cross-channel operator of spin $\ell$ were obtained explicitly by decomposing the crossing kernels into Wilson functions. 
The results involve $(2\ell+1)$ Wilson functions 
with coefficients given by double finite sums of a product of two ${}_4F_3$ hypergeometric series. 
We checked at low $\ell$ that they are equivalent to the $\log z$ part of \eqref{general-LI}, up to normalizations. 
Note that at low direct-channel spin the formula for $\ell>0$ in \cite{Sleight:2018ryu} has singularities and requires additional manipulations to obtain sensible results. 
Our formula \eqref{general-LI} is much simpler and more efficient, which also includes the regular part associated with the corrections to OPE coefficients.  }

The Lorentzian inversion integral should commute with the $\a$-derivative when no singularity is encountered. 
For efficient numerical evaluation, 
one can write the well-poised ${}_7F_6$ as a sum of two 1-balanced ${}_4F_3$ functions, 
replace $\pa_\a f(\a)$ by $[f(\a)-f(0)]/\a$, and set $\a$ to a small number. 
For conserved currents, one should set $\a\ll\g\ll1$ or use \eqref{conserved} to derive the inversion. 
Our formula \eqref{general-LI} can efficiently reproduce the 3d nonperturbative results in \cite{Albayrak:2019gnz} 
from dimensional reduction \cite{Hogervorst:2016hal}. 
\section{Conclusion}
In summary, we presented a new closed-form expression \eqref{closed-form} for the lightcone limit of 4-point conformal blocks  in the cross-channel, 
where the external operators are generic scalar primaries. 
This compact result was based on the basis functions \eqref{basis-fun} inspired by the nonperturbative pole structure in spin of the Lorentzian inversion. 
Our expression applies to intermediate operators of arbitrary spin in general dimensions. 
When the external scalars have identical scaling dimensions, 
we also provided the expression \eqref{general-LI} for 
the Lorentzian inversion of a general cross-channel conformal block, 
which will be particularly useful for the investigations with $d\neq 2, 4$, 
as general compact expressions were not available in the past. 
They include the analytic conformal bootstrap of 
the Wilson-Fisher fixed points \cite{Alday:2017zzv,Henriksson:2018myn}, conformal field theories in 
$d=3$ dimensions \cite{Aharony:2018npf, Albayrak:2019gnz} 
and $d>4$ dimensions \cite{Alday:2016jfr}. 

It would be interesting to better understand \eqref{closed-form}, 
such as the connections to higher order differential equations \cite{Hogervorst:2013kva, Caron-Huot:2017vep} 
and integrability \cite{Isachenkov:2016gim, Isachenkov:2017qgn}.  
Using Casimir equations, one can compute the subleading terms in the lightcone expansion. 
It turns out that several nice features of the leading term \eqref{closed-form} are shared by the subleading terms to all orders, 
so our results also lead to a new analytical understanding of the complete conformal blocks \cite{Li:2019cwm}.
For simplicity, we focused on the Lorentzian inversion with identical external dimensions, 
but one can extend \eqref{general-LI} to the generic case using \eqref{closed-form}.  

In the original paper \cite{Caron-Huot:2017vep}, it was proposed that the Lorentzian inversion formula applies to direct-channel operators of spin $J>1$ in a unitary theory, 
where the Regge limit is well-behaved. 
Recently, it has been noticed that analyticity in spin can be extended to $J=0$ \cite{Turiaci:2018dht}\cite{Alday:2017zzv,Henriksson:2018myn}. 
Here, the poles with negative spin were instrumental in determining the basis functions \eqref{basis-fun}, 
which are reminiscent of the nonunitary poles in scaling dimension 
in the Zamolodchikov-like recursion relations \cite{Kos:2013tga, Penedones:2015aga}. 
Along these lines, we may eventually apply the Lorentzian inversion formula to nonunitary CFTs 
after a more careful treatment of analytic continuation. 
Due to absence of reflection positivity, 
many statistical physics models are related to nonunitary CFTs, 
including the Wilson-Fisher fixed points \cite{Hogervorst:2015akt}.  
We may study them by combining the analytical toolkit with the OPE truncation methods \cite{Gliozzi:2013ysa, Gliozzi:2014jsa, Li:2017ukc}, 
as proposed in \cite{Simmons-Duffin:2016wlq}. 

\begin{acknowledgments}
I am grateful to Shinobu Hikami, Junchen Rong, Hidehiko Shimada and Ning Su for inspiring discussions. 
This work was supported by Okinawa Institute of Science and Technology Graduate University (OIST) 
and JSPS Grant-in-Aid for Early-Career Scientists (KAKENHI No. 19K14621).
\end{acknowledgments}

\end{document}